\documentclass[iop]{emulateapj}
\usepackage{color}

\usepackage{ulem}
\usepackage{amsmath}
\usepackage{enumerate}
\usepackage[colorlinks=true,citecolor=blue,breaklinks=true]{hyperref}

\newcommand\ergs{erg~s$^{-1}$}
\newcommand\ergcms{erg~cm$^{-2}$~s$^{-1}$}

\newcommand\fnunu{erg~cm$^{-2}$~s$^{-1}$~Hz$^{-1}$}
\newcommand\fnu{erg~s$^{-1}$~Hz$^{-1}$}

\newcommand{\angstrom}{\text{\normalfont\AA}}

\newcommand{\hbeta}{H{$\beta$}}

\newcommand{\halpha}{H{$\alpha$}}

\newcommand{\CaIIk}{Ca\,{\sc ii}\,K\,$\lambda$3933}
\newcommand{\CaIIh}{Ca\,{\sc ii}\,H\,$\lambda$3968}

\newcommand{\MgII}{Mg\,{\sc ii}}

\newcommand{\OIII}{[O\,{\sc iii}]}

\newcommand{\OIIIb}{[O\,{\sc iii}]\,$\lambda$5007}
\newcommand{\OIIIab}{[O\,{\sc iii}]\,$\lambda\lambda$4959,5007}
\newcommand{\FeII}{Fe\,{\sc ii}}

\newcommand{\OII}{[O\,{\sc ii}]}
\newcommand{\OIIwave}{[O\,{\sc ii}]\,$\lambda$3728}

\newcommand{\NII}{[N\,{\sc ii}]}
\newcommand{\NIIwave}{[N\,{\sc ii}]\,$\lambda$6583}
\newcommand{\SII}{[S\,{\sc ii}]}

\begin{document}

\title{PHL 6625: A Minor Merger-Associated QSO Behind NGC 247}

\author{Lian Tao\altaffilmark{1,2,3},  Hua Feng\altaffilmark{1,2}, Yue Shen\altaffilmark{4}, Luis C. Ho\altaffilmark{5,6}, Junqiang Ge\altaffilmark{7}, Philip Kaaret\altaffilmark{8}, Shude Mao\altaffilmark{9,2,7,10}, Xin Liu\altaffilmark{4}}

\altaffiltext{1}{Department of Engineering Physics, Tsinghua University, Beijing 100084, China}
\altaffiltext{2}{Center for Astrophysics, Tsinghua University, Beijing 100084, China}
\altaffiltext{3}{Cahill Center for Astronomy and Astrophysics, California Institute of Technology, Pasadena, CA 91125, USA}

\altaffiltext{4}{Department of Astronomy, University of Illinois at Urbana-Champaign, Urbana, IL 61801, USA}

\altaffiltext{5}{Kavli Institute for Astronomy and Astrophysics, Peking University, Beijing 100087, China}
\altaffiltext{6}{Department of Astronomy, Peking University, Beijing 100087, China}

\altaffiltext{7}{National Astronomical Observatories, Chinese Academy of Sciences, 20A Datun Road, Chaoyang District, Beijing 100012, China}

\altaffiltext{8}{Department of Physics and Astronomy, University of Iowa, Iowa City, IA 52242, USA}

\altaffiltext{9}{Department of Physics, Tsinghua University, Beijing 100084, China}

\altaffiltext{10}{Jodrell Bank Centre for Astrophysics, School of Physics and Astronomy, The University of Manchester, Oxford Road, Manchester M13 9PL, UK}

\shorttitle{PHL 6625: QSO associated with a minor merger}
\shortauthors{Tao et al.}

\begin{abstract}
PHL 6625 is a luminous quasi-stellar object (QSO) at $z = 0.3954$ located behind the nearby galaxy NGC 247 ($z = 0.0005$). Hubble Space Telescope (HST) observations revealed an arc structure associated with it. We report on spectroscopic observations with the Very Large Telescope (VLT) and multiwavelength observations from the radio to the X-ray band for the system, suggesting that PHL 6625 and the arc are a close pair of merging galaxies, instead of a strong gravitational lens system. The QSO host galaxy is estimated to be $(4-28) \times 10^{10}$~$M_\sun$, and the mass of the companion galaxy of is estimated to be $M_\ast = (6.8 \pm 2.4) \times 10^{9}$~$M_\sun$, suggesting that this is a minor merger system. The QSO displays typical broad emission lines, from which a black hole mass of about $(2-5) \times 10^8$~$M_\sun$ and an Eddington ratio of about 0.01--0.05 can be inferred. The system represents an interesting and rare case where a QSO is associated with an ongoing minor merger, analogous to Arp 142. 
\end{abstract}


\section{Introduction}
\label{sec:intro}

Quasi-stellar objects (QSOs) or quasars are believed to be powered by accretion onto supermassive black holes in the centers of galaxies. How quasars are triggered is still not clear and under investigation. It is generally accepted that major mergers can trigger substantial star formation and possibly accretion onto the central black hole. \citet{Sanders1988} proposed that major mergers, particularly between gas-rich disk galaxies, might drive gas to flow toward the nuclear region and initiate starburst, and then the triggering of the quasar phase. This picture is in good agreement with numerical simulations \citep[e.g.,][]{Hernquist1989,Hopkins2006}. Observations of the ultraluminous infrared galaxies (ULIRGs) indicate that merging features \citep{Sanders1996} and the quasar fraction \citep{Kartaltepe2010} are strongly correlated with their IR luminosities, implying that major mergers and quasar activity may have a connection.

Minor mergers have been proposed to induce the fueling of low-luminosity AGNs and explain some observational features, such as the random orientation of narrow-line regions with respect to the host disks, the excess of ring-like structures and their off-center locations, and their amorphous morphology \citep[e.g.,][]{Taniguchi1999,Combes2009}. Several authors suggested separating the fueling mechanisms for quasars from that for low-luminosity AGNs: major mergers trigger quasars, while minor mergers trigger low-luminosity AGNs \citep{Hopkins2009,Taniguchi2013}. However, it is still uncertain if minor mergers could trigger high-luminosity AGN (quasar) activity. Significant fine structures such as shells and tidal tails were observed in deep Hubble Space Telescope (HST) images of four out of five elliptical low-redshift quasar host galaxies, which can be explained as due to minor mergers between a dwarf galaxy and a giant elliptical galaxy \citep{Bennert2008}. Thus, \citet{Bennert2008} suggested that minor mergers might trigger the observed quasar activity. Moreover, \citet{Tadhunter2014} examined 32 quasar-like AGN host galaxies and found that their dust masses were intermediate between those of quiescent elliptical galaxies and ULIRGs, suggesting that most of these AGNs were triggered in mergers between giant elliptical galaxies and relatively low gas mass companion galaxies.

It is also debatable whether the central black holes can be ignited in an interacting close pair at the early stage of merging when they start to have tidal interactions but are still spatially separate; some observations lead to a positive answer \citep[e.g.,][]{Ellison2011,Silverman2011,Liu2012} while others do not \citep[e.g.,][]{Ellison2008}. 

This study is unable to address these questions from a statistical point of view, but it presents an interesting case where a luminous QSO (PHL 6625) is found in a close pair of a merging system in the local universe. PHL 6625 ($z = 0.3954$, see~\S~\ref{sec:vlt_spec} for details) is a radio-quiet QSO projected on the outskirts of a nearby spiral galaxy, NGC 247 \citep[$z = 0.0005$,][]{Karachentsev2013}. It was detected as a redshifted object behind NGC 247 by \citet{Margon1985}. In 2011 October, \citet{Tao2012} serendipitously discovered that PHL 6625 was associated with an arc structure on an HST image. We thus conducted new observations to further investigate its nature.

\begin{deluxetable*}{cllccl}
\tablecolumns{6}
\tablewidth{\textwidth}
\tablecaption{GALFIT modeling of the QSO image
\label{tab:galfit}}
\tablehead{
\colhead{Model} & \colhead{QSO (model)} & \colhead{Host (model)} & \colhead{QSO (mag)}  & \colhead{Host (mag)} & \colhead{Reduced $\chi^2$ (dof)} \\
\colhead{(1)} & \colhead{(2)} & \colhead{(3)} & \colhead{(4)}  & \colhead{(5)} & \colhead{(6)}
}
\startdata
1a  & PSF            & S{\'e}rsic ($n=0.6$)            & $-22.01$ & $-21.15 $ & 1.803 (31954) \\
1b & PSF + S{\'e}rsic ($n=0.3$) & S{\'e}rsic ($n=8.3$)           & $-22.28$ & $-20.58 $ & 1.318 (31947) \\
2a & PSF            & Exponential            & $-22.42$ & $-19.32 $ & 2.402 (31955) \\
2b & PSF + S{\'e}rsic ($n=0.2$) & Exponential           & $-22.30$ & $-19.52 $ & 1.393 (31948) \\
3a & PSF           & S{\'e}rsic ($n=0.3$) + S{\'e}rsic ($n=8.3$) & $-21.87$ & $-21.62 $ & 1.318 (31947) \\
3b & PSF + S{\'e}rsic ($n=0.3$) & S{\'e}rsic ($n=1.0$) + S{\'e}rsic ($n=8.4$) & $-22.28$ & $-20.59 $ & 1.316 (31940) \\
4a & PSF            & S{\'e}rsic ($n=0.2$) + Exponential & $-21.90$ & $-21.33 $ & 1.393 (31948) \\
4b & PSF + S{\'e}rsic ($n=0.3$) & S{\'e}rsic ($n=8.3$) + Exponential & $-22.28$ & $-20.59 $ & 1.316 (31941)
\enddata
\tablecomments{Some models are identical (1b = 3a, 2b = 4a); the difference is whether the small-index S{\'e}rsic is attributed to the QSO or the host galaxy. 
}
\end{deluxetable*}

The projected distance of the quasar from NGC 247 at the distance of NGC 247 \citep[3.4~Mpc,][]{Gieren2009} is about 4.4~kpc. Thanks to its spatial location and strong X-ray emission, PHL 6625 is of interest in probing the halo of NGC 247 and acts as a useful probe to detect the ``missing baryons,'' which have an observational deficit \citep[e.g.,][]{Shull2012} compared with cosmological predictions. Some of the missing baryons in the local universe are thought to be locked in the warm-hot intergalactic medium, which can be detected via X-ray absorption lines in the presence of a background QSO using next generation telescopes \citep{Yao2012}.

In this paper, we report spectroscopic observations for PHL 6625 and its nearby arc structure with the Very Large Telescope (VLT) of the European Southern Observatory (ESO), along with multiwavelength observations from the radio to the X-ray band. We adopt a cosmology with $h = 0.7$, $\Omega_m = 0.3$, and $\Omega_\Lambda = 0.7$ and a luminosity distance of 2.14 Gpc to PHL 6625 ($z = 0.3954$).

\section{Observations and data analysis}
\label{sec:obs}

\subsection{\it HST imaging}

\begin{figure}
\centering
\includegraphics[width=1.0\columnwidth]{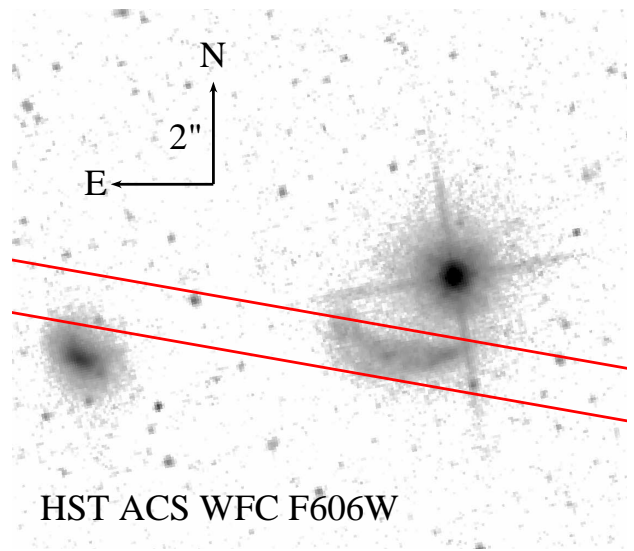}
\caption{HST image around PHL 6625. The red lines indicate the VLT slit positions for the arc structure. The three objects from east to west are an anonymous galaxy (not NGC 247), the arc structure, and the QSO PHL 6625, respectively. The arrows have a length of 2\arcsec\ (1\arcsec\ = 5.4~kpc at a redshift of 0.3970 assuming $h = 0.7$, $\Omega_m = 0.3$, and $\Omega_\Lambda = 0.7$). 
\label{fig:hst}}
\end{figure}

HST observed the northwestern region of NGC 247 on 2011 October 11 (proposal ID 12375), using the broadband filter F606W of the Wide Field Channel (WFC) on the Advanced Camera for Surveys (ACS) with two sub-exposures for a total exposure of 846~s. The observation was designed to use the QSO PHL~6625, which was known to be bright in both the X-ray and optical bands, to align the Chandra and HST images to improve their relative astrometry \citep{Tao2012}. However, it serendipitously found that the QSO was associated with an arc structure to its southeast; see Figure~\ref{fig:hst}. The QSO is projected at about 0.4 times the $R_{25}$ radius of NGC 247 \citep{de Vaucouleurs1991} in a relatively uncrowded environment. 

\begin{figure}
\centering
\includegraphics[width=\columnwidth]{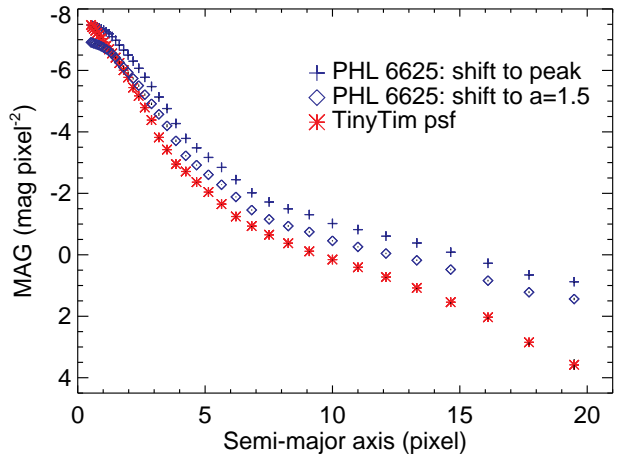}
\caption{Azimuthally averaged radial profiles of PHL 6625 and the TinyTim PSF. MAG is the mean isophotal magnitude. Crosses show the profile of PHL 6625 shifted to match the peak of the TinyTim PSF and diamonds show the PHL 6625 profile shifted to match the TinyTim PSF at a radius of 1.5 pixels. In either case, the profile of PHL 6625 is significantly broader than that of the PSF, indicating that there is a host galaxy component.
\label{fig:profile}}
\end{figure} 

A reasonable point-spread function (PSF) is needed to analyze the QSO image. Looking through the archival HST images\footnote{The observing date of proposal ID 12375 was later than the Servicing Mission 4 (SM4) of HST. In SM4, the ACS was repaired, and the PSF might be affected. Therefore, the PSF stars are searched in the observations after SM4.}, isolated bright stars with a flat background cannot be found at a similar chip position. We thus derived a synthetic PSF by synthesizing PSF models generated by the TinyTim tool \citep{Krist1995} onto the flat-fielded calibrated (\_flc) data and drizzling them into science images using the {\tt astrodrizzle} task. The azimuthally averaged radial profiles of the QSO and the TinyTim PSF are computed using the {\tt ellipse} task in the IRAF/STSDAS package and plotted in Figure~\ref{fig:profile} for comparison. They are shifted to have the same brightness at the center (0.5 pixel). As the core ($3\times3$ pixels) of the QSO image is saturated due to high brightness and may be not useful, we also compared the two profiles by matching their magnitudes at a radius of 1.5 pixels, beyond which there is no saturation. In either way, the PSF profile is significantly narrower than that for the QSO, suggestive of an additional component (likely the host galaxy) underneath the QSO component.

\begin{figure*}
\centering
\includegraphics[width=\textwidth]{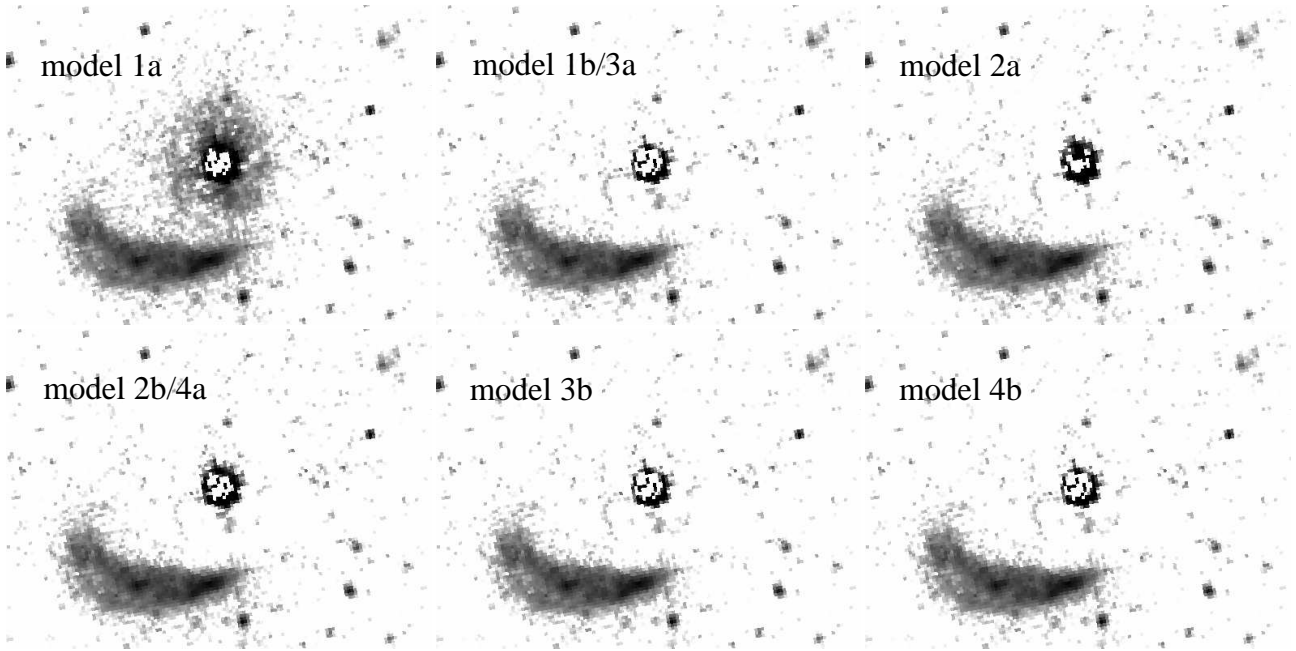}
\caption{GALFIT fit residuals for the models listed in Table~\ref{tab:galfit}.
\label{fig:galfit}}
\end{figure*}

The QSO image is decomposed into a QSO component and a host galaxy component using GALFIT \citep{Peng2002,Peng2010}. The arc structure, central saturated $3 \times 3$ pixels, and nearby stars are masked away during the fit. Without information from the central pixels,  it is hard to reconstruct the bulge component unless it is sufficiently extended. We thereby experimented with several models to explore the systematics in the modeling. For the QSO, we also tried a PSF with the addition of a small-index S{\'e}rsic profile to account for PSF artifacts. For the host galaxy, we tested with either a single-component model (a S{\'e}rsic or an exponential disk) or a two-component model (two S{\'e}rsics or a S{\'e}rsic + an exponential disk). These lead to eight combinations of models, tabulated in Table~\ref{tab:galfit}. The simplest models with only two components (model 1a and 2a) do not provide adequate fits, while any model with three or four components can fit the image similarly well. The residuals are shown in Figure~\ref{fig:galfit} for comparison. Therefore, we discard the simplest (two-component) models and adopt the magnitude range derived from other models as a conservative estimate of its uncertainty, for both the QSO and the host galaxy.

The flux of the companion arc galaxy is measured using aperture photometry, with a visually defined polygon aperture and nearby source-free regions for background estimate. Assuming a flat spectrum ($F_\lambda \propto \lambda^{0}$ ) for the arc and the QSO host galaxy, and a power-law spectrum for the QSO (see ~\S~\ref{sec:qso_gfit} for details), the measured count rates can be translated to dereddened, K-corrected, absolute $B$ magnitudes of [$-$21.87, $-$22.30],  [$-$19.52, $-$21.62], and $-$19.86, respectively, for the QSO, the QSO host galaxy, and the arc galaxy. Assuming a solar $B$-band magnitude of 5.48 \citep{Binney1998}, the total luminosity in the $B$-band is, respectively, $(0.87-1.30) \times 10^{11}$, $(1.00-6.93) \times 10^{10}$, and $1.37 \times 10^{10}$ $L_\sun$ for the three objects in the same order. The projected size of the arc galaxy is roughly 16 by 4 kpc.

\subsection{\it VLT spectroscopy}
\label{sec:vlt_spec}

To unveil the nature of the QSO and the arc galaxy, we conducted spectroscopic observations with the ESO 8.2-m diameter VLT at Paranal in Chile (program ID 091.A-0149(A)), using the FORS2 long-slit spectrograph mounted on the unit telescope 1 (UT1). A red (GRIS\_300I) and blue (GRIS\_600B) grism is used, respectively, to cover a wavelength range from $\sim$3700--10000\AA. The blue setup of the arc is spilt into two identical observations. Each observation consists of two (for the QSO) or three (for the arc) observation blocks (OBs) with an offset of 3\arcsec\ along the spatial direction between successive ones for bad pixel and cosmic-ray removal. We used a 1\arcsec\ slit and a $2 \times 2$ binning of pixels, resulting in a sampling of 2.8\AA/pixel in the red and 1.32\AA/pixel in the blue. The spectral resolution in FWHM found from the lamp lines varies from 11.3--12.9\AA\ in the red and 5.3--5.9\AA\ in the blue. For the QSO observations, the slit is oriented across both the QSO and the central region of the arc, while for the arc observations, the slit is placed along its major axis. The observational log is listed in Table~\ref{tab:vlt} and the location of the slit for the arc observation is displayed in Figure~\ref{fig:hst}.

\tabletypesize{\scriptsize}
\begin{deluxetable}{llllll}
\tablecolumns{6}
\tablewidth{\columnwidth}
\tablecaption{Log of {\it VLT} FORS2 observations
\label{tab:vlt}}
\tablehead{
\colhead{Obj} & \colhead{Start time} & \colhead{Grism} & \colhead{Exposure}  & \colhead{Airmass} & \colhead{Seeing} \\
\colhead{} & \colhead{ (2013; UT)} & \colhead{} & \colhead{(s)}  & \colhead{} & \colhead{(\arcsec)}
}
\startdata
QSO & Jul 17 09:04:08 & 300I & $150 \times 2$ & 1.009 & 0.6 \\
QSO & Jul 18 09:30:59 & 600B & $150 \times 2$ & 1.003 & 0.7 \\
arc & Aug 04 05:54:06 & 300I & $940 \times 3$ & 1.169 & 0.7 \\
arc & Aug 04 06:51:06 & 600B & $810 \times 3$ & 1.054 & 0.9 \\
arc & Aug 04 07:44:30 & 600B & $810 \times 3$ & 1.012 & 0.7
\enddata
\end{deluxetable}

\begin{figure*}
\centering
\includegraphics[width=\textwidth]{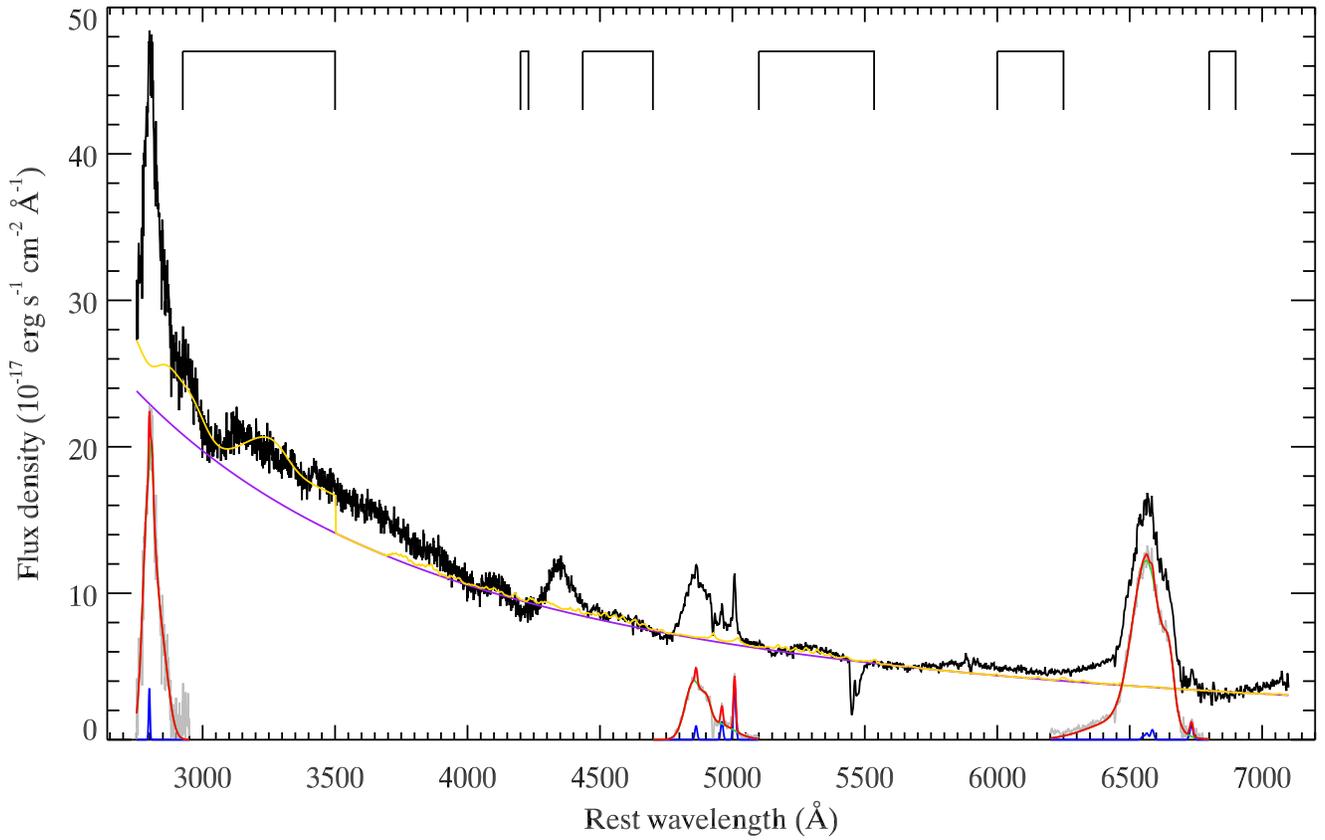}
\caption{VLT FORS2 spectrum of the QSO PHL 6625 with model components. We fitted the spectrum from 2750 to 7100\AA, where the S/N is high. The windows used to fit the continuum are marked in black bars, and pixels with telluric absorptions around 5450\AA\ were excluded in the fitting. The dereddened spectrum is shifted to the rest frame of the QSO at $z = 0.3954$.
{\it Purple}: power law; 
{\it yellow}: pseudo-continuum = power law + \FeII\ templates; 
{\it red}: summed line model; 
{\it blue}: narrow-line components; 
{\it green}: broad-line components.
\label{fig:qso_spec}}
\end{figure*}

The {\tt esorex} package was used to create bias-subtracted, flat-fielded, and wavelength-calibrated 2D spectra for each OB, using calibration files obtained in the same night. The {\tt imcombine} task in IRAF was then used to combine different OBs and remove cosmic rays with the option {\tt crreject}. The 1D spectra were extracted using the {\tt apall} task. The trace information was obtained by fitting the QSO spectra and was applied for the extraction of the arc spectra. The source aperture size is around 10--11 pixels for the QSO and 18--20 pixels for the arc, and the background was estimated by fitting fluxes from two source-free regions on each side. The standard star LTT 1020 observed in the same night as the arc observations was used for flux calibration. An extinction table created on 2011 January 18 was used for atmospheric extinction correction.

The red or blue spectra from individual OBs for either the QSO or arc are averaged. The red spectra have a flux higher than the blue spectra in their overlapping region, which is due to a smaller PSF image size in the red. We thus scale the blue flux by a constant factor of $\sim$1.2 to yield a consistent flux in their overlapping wavelength range. Spectral smoothing is done with a 5 pixel median filter.

\begin{figure*}
\centering
\includegraphics[width=\textwidth]{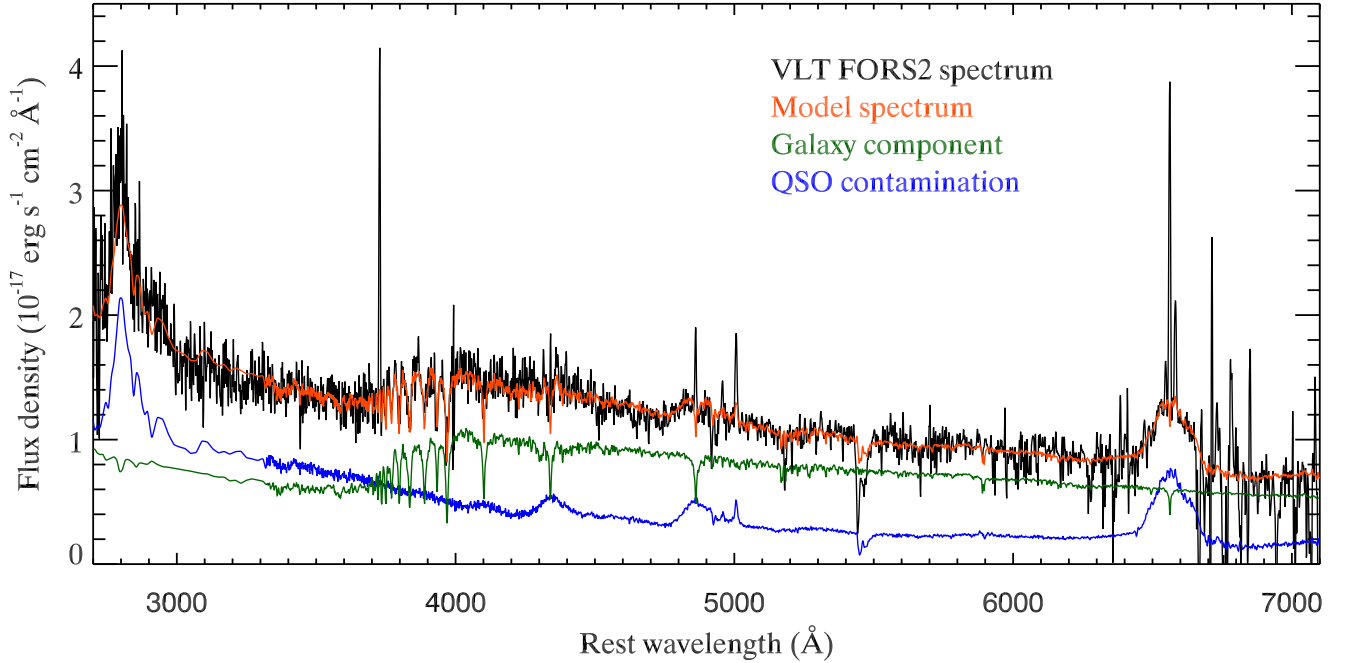}
\caption{VLT FORS2 dereddened spectrum (black) of the arc galaxy with the QSO contamination (blue), the galaxy population synthesis model (green) and the total model (red). All emission lines and telluric absorptions are masked during the population synthesis fit. The dereddened spectrum is shifted to the rest frame of the arc at $z = 0.3970$.
\label{fig:arc_spec}}
\end{figure*}

The observed QSO and arc spectra are shown in Figures~\ref{fig:qso_spec} and \ref{fig:arc_spec}, respectively. The QSO spectrum exhibits characteristic emission lines, such as broad \MgII, \hbeta\ and \halpha, narrow \OIIIab\ and Balmer lines. The absorption features around 5450\AA\ are telluric. The arc spectrum shows some narrow emission lines, such as \OIIwave, \hbeta, \OIIIb, \halpha\ and \NIIwave\ and some weak absorption lines, such as \CaIIk\ and \CaIIh. The arc spectrum is contaminated by the QSO, showing broad \halpha\ and \MgII\ lines (see ~\S~\ref{sec:arc_fit} for details). The redshift of the QSO is measured to be $0.3954\pm0.0001$ from the \OIIIab\ emission lines. The redshift of the arc, using the QSO-contamination-subtracted spectrum, is measured to be $0.3970\pm0.0001$ via \OII\ and \hbeta\ emission lines, and consistent with the value of $0.3968\pm0.0003$ measured via the \CaIIk\ and \CaIIh\ absorption lines. The arc is likely located nearer than the QSO and moving toward it with a line-of-sight velocity of roughly 340 km~s$^{-1}$.

\subsubsection{Decomposition of the QSO spectrum}
\label{sec:qso_gfit}

The QSO spectrum is decomposed into multiple emission components following \citet{Shen2008}. The fitting is performed in the rest-frame wavelength range of 2750--7100\AA, where the S/N is sufficiently high. First, a pseudo-continuum was fitted to the spectrum in some continuum windows\footnote{2925--3500\AA, 4200--4230\AA, 4435--4700\AA, 5100--5535\AA, 6000--6250\AA, and 6800--6900\AA\ at the rest frame.}, consisting of a power-law component and \FeII\ templates in both the \MgII\ region \citep{Salviander2007} and the \hbeta\ region \citep{Boroson1992}. The continuum-subtracted line spectrum was then fitted with multiple Gaussian components: three Gaussians for each of the \halpha, \hbeta, and \MgII\ broad components; five Gaussians for narrow lines in the \halpha\ region\footnote{Two for \SII, two for \NII, and one for narrow \halpha.}, three near \hbeta\footnote{Two for \OIII\ and one for narrow \hbeta.}, and one for narrow \MgII. All of the narrow lines are imposed to have the same shift and width. 

\begin{deluxetable*}{cll}
\tablecolumns{3}
\tablewidth{0pc}
\tablecaption{Global fitting results for the QSO optical spectrum
\label{tab:gfit}}
\tablehead{
\colhead{} & \colhead{} & \colhead{Note}
}
\startdata
$\alpha$ & $-2.171 \pm 0.010$ & Power-law spectral index, $f_\lambda \propto \lambda^\alpha$\\
$L_{3000}$ & $4.54 \times 10^{44}$~\ergs & $\lambda L_\lambda(3000\angstrom)$ of the power-law component \\
$L_{5100}$ & $2.44 \times 10^{44}$~\ergs & $\lambda L_\lambda(5100\angstrom)$ of the power-law component \\
FWHM(\MgII)   & 6269~km~s$^{-1}$ & Broad component \\
FWHM(\hbeta)  & 6447~km~s$^{-1}$ & Broad component \\
FWHM(\halpha) & 7314~km~s$^{-1}$ & Broad component \\
$\log (M/M_\sun)$ & 8.43 & Based on $L_{5100}$ and FWHM(\hbeta), $\alpha = 0.5$ and $\beta = 2$ (Ref.\ 1) \\
$\log (M/M_\sun)$ & 8.74 & Based on $L_{5100}$ and FWHM(\hbeta), $\alpha = 0.533$ and $\beta = 2$ (Ref.\ 2) \\
$\log (M/M_\sun)$ & 8.47 & Based on $L_{5100}$ and FWHM(\hbeta), $\alpha = 0.5$ and $\beta = 1.09$ (Ref.\ 3) \\
$\log (M/M_\sun)$ & 8.29 & Based on $L_{5100}$ and FWHM(\hbeta), $\alpha = 0.572$ and $\beta = 1.200$ (Ref.\ 1) \\
$\log (M/M_\sun)$ & 8.25 & Based on filtered luminosities (Ref. 1)
\enddata

\tablerefs{(1): \citet{Feng2014}; (2): \citet{Ho2015}; (3): \citet{Wang2009}}

\end{deluxetable*}

The monochromatic continuum luminosity $\lambda L_\lambda$ of the power-law component at the rest frame 3000\AA\ and 5100\AA, the power-law spectral index, and the broad-line width derived from the global fitting are listed in Table~\ref{tab:gfit}. Based on the radius-luminosity ($R-L$) relation, the black hole mass in the QSO can be estimated from the line width and the continuum luminosity with a single-epoch spectrum, i.e., $M_{\rm BH} \propto L^\alpha {\rm FWHM}^\beta$ \citep{Shen2013}. Here we use five recipes that were calibrated against \hbeta\ reverberation-mapped masses: the updated \citet{Vestergaard2006} formula described in \citet{Feng2014} that assumes theoretical slopes ($\alpha = 0.5$ and $\beta = 2$) on the luminosity and line width, the calibration of \citet{Ho2015} based on a best-fit slope for the $L-R$ relation \citep[$\alpha = 0.533$;][]{Bentz2013} and $\beta = 2$, the calibration of \citet{Wang2009} with $\alpha = 0.5$ and a best-fit slope for the single-epoch FWHM versus rms line dispersion relation ($\beta = 1.09$), one based on best-fit slopes for both the luminosity and line width \citep[$\alpha = 0.572$ and $\beta = 1.200$;][]{Feng2014}, and a novel technique \citep{Feng2014} that establishes a correlation between the black hole mass and filtered luminosities (luminosities extracted in two wavelength bands). All the recipes give consistent results for a black hole mass of about $(2-5) \times 10^8$~$M_\sun$, also listed in Table~\ref{tab:gfit}. If we use the second moment \citep[line dispersion;][]{Peterson2004} instead of the FWHM, the inferred black hole mass is consistent with the result above within the intrinsic scatter (a factor of $\sim$2).

\begin{deluxetable*}{cccccc}
\tablecolumns{6}
\tablewidth{0pc}
\tablecaption{Stellar population synthesis results for the Arc galaxy
\label{tab:pop}}
\tablehead{
\colhead{$M_\ast / M_\sun$}  & \colhead{Extinction ($A_V$)}  & \colhead{$<\log [{\rm Age}_L~(\rm yr)]>$}  & \colhead{$<\log [{\rm Age}_M~(\rm yr)]>$} & \colhead{$<Z_L/Z_{\odot}>$} & \colhead{$<Z_M/Z_{\odot}>$}
}
\startdata
$(6.8 \pm 2.4) \times 10^9$ & $0.24 \pm 0.06$ & $8.44 \pm 0.16$ & $9.8 \pm 0.2$ & $0.08 \pm0.02$ & $0.7 \pm 0.5$
\enddata
\end{deluxetable*}

\tabletypesize{\scriptsize}
\begin{deluxetable}{ccccc}
\tablecolumns{5}
\tablewidth{\columnwidth}
\tablecaption{Emission line dereddened luminosities of the Arc galaxy
\label{tab:emi}}
\tablehead{
\colhead{\OIIwave} & \colhead{\hbeta} & \colhead{\OIIIb} & \colhead{\halpha} & \colhead{\NIIwave} 
}
\startdata
$13.6 \pm 2.7$ & $4.1 \pm 0.7$ & $3.6 \pm 0.6$ & $13.2 \pm 1.4$ & $5.0 \pm 0.6$
\enddata
\tablecomments{in units of $10^{40}$~\ergs}
\end{deluxetable}

\begin{figure}
\centering
\includegraphics[width=\columnwidth]{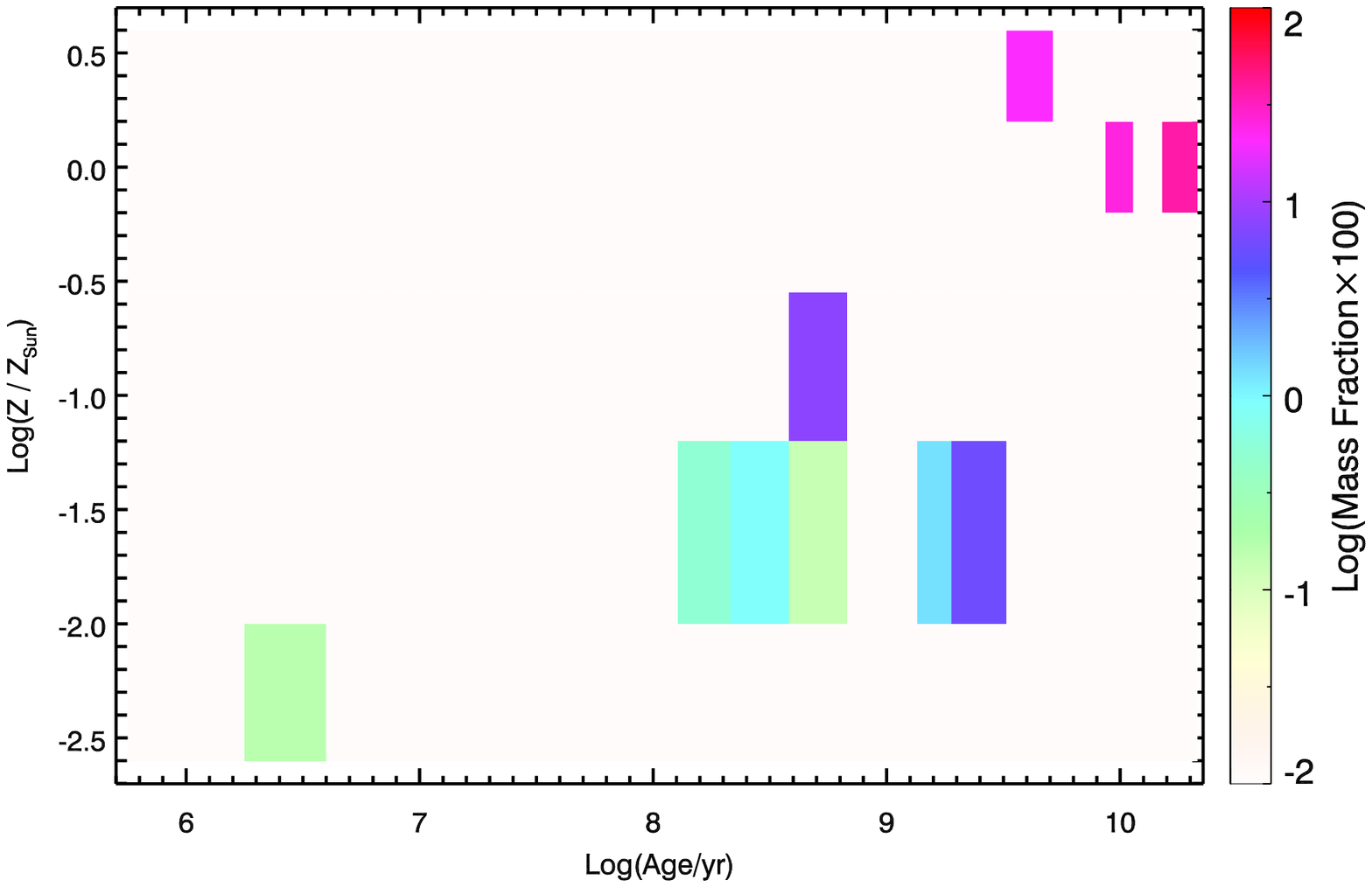}
\caption{Decomposed star formation history of the arc galaxy.
\label{fig:arc_sfh}}
\end{figure}

\subsubsection{Population synthesis for the Arc galaxy}
\label{sec:arc_fit}

The stellar population synthesis code STARLIGHT \citep{CidFernandes2005} was used to fit the arc spectrum. As the arc spectra may be contaminated by flux from the QSO, we added the QSO spectrum into the simple stellar population (SSP) fitting library as a model template. We then fit the arc spectrum with the stellar components using the BC03 theoretical library \citep[150 SSPs with 25 ages and 6 metallicities,][]{Bruzual2003} with Chabrier's initial mass function \citep{Chabrier2003}. Assuming that the flux error follows a Gaussian distribution, we generated 50 mock spectra to estimate the parameter uncertainties.

The scaling factor for a QSO contribution to the arc spectrum is found to be 3.5\% from the fit. We fit a Moffat function to the QSO on the acquisition image, and estimate that the QSO roughly contributes 3.4\% of its flux to the arc aperture, in reasonable agreement with the result from the spectral fitting. After the QSO contribution is removed, the arc spectrum barely shows broad-line components (\halpha\ and \MgII).

The STARLIGHT fitting results are listed in Table~\ref{tab:pop}, including the stellar mass, star-derived extinction, and luminosity-weighted and mass-weighted ages and metallicities. The best-fit model spectrum is shown in Figure~\ref{fig:arc_spec} and the decomposed star formation history (SFH) is shown in Figure~\ref{fig:arc_sfh}. The emission line luminosities are measured from the galaxy model and QSO template subtracted spectrum. The gas-derived extinction was derived from the star-derived extinction, assuming the ratio between the gas-derived and star-derived extinction to be 0.44 \citep{Calzetti2001}. The dereddened luminosities are listed in Table~\ref{tab:emi}. The ratio of   ${\rm H}\alpha / {\rm H}\beta$, $3.2\pm0.6$, is consistent with the Balmer decrement \citep[2.86,][]{Osterbrock1989}. The luminosity of the \OIIIb\ line only accounts for 0.1\% of the observed flux in the HST F606W filter, suggesting that the arc is truly made of stars rather than some extended ionized gas. 

A single Gaussian component is able to fit each of the narrow lines in the contamination-subtracted arc spectrum, and no obvious residuals are seen. The observed FWHM of the emission/absorption lines corrected for instrumental broadening is consistent with zero within errors, which agrees with the result that the stellar velocity dispersion ($\sim$200~km~s$^{-1}$) derived from the fundamental plane \citep[summarized in][]{Kormendy2013} is smaller than the instrument dispersion (> 300~km~s$^{-1}$) and hence unresolved.

The star formation rate (SFR) is estimated from the strongest emission line \halpha\ \citep{Kennicutt1994,Madau1998,Kennicutt1998} assuming solar abundance and \citet{Salpeter1955}'s IMF,

\begin{equation}
\label{eq:SFR}
{\rm SFR} = 7.9 \times 10^{-42} \; \frac{\; L({\rm H}\alpha)\; }{{\rm erg \; s}^{-1}} \quad {M_\sun \; {\rm yr}^{-1}}.
\end{equation} 

We derived an SFR for the arc galaxy, ${\rm SFR} = (1.04 \pm 0.11)$~$M_\sun$~yr$^{-1}$, and a specific SFR (SFR per stellar-mass unit), ${\rm sSFR} = {\rm SFR} / M_\ast$ = 0.15~Gyr$^{-1}$. Using \citet{Kennicutt1998}'s calibration, the SFR derived from the \OIIwave\ line gives a marginally consistent result, $(1.9 \pm 0.7)$~$M_\sun$~yr$^{-1}$. Given a redshift of 0.3954 and the ${\rm SFR}$ range estimated above, the stellar mass is estimated to be $(1.7-7.9) \times 10^{9}$~$M_\sun$ if the source lies on the main sequence of star-forming galaxies \citep{Whitaker2012}, consistent with $M_\ast = (6.8\pm2.4) \times 10^{9}$~$M_\sun$ derived from the population synthesis.

Given the metallicities and ages in Table~\ref{tab:pop}, we can predict a mass-to-light ratio ($M_\ast/L_{\rm B}$) of $\sim0.1-0.2$ and $\sim1.3-5.0$ in the $B$-band following \citet{Maraston2005}, for luminosity-weighted and mass-weighted measurements, respectively. Using the mass obtained from the stellar population synthesis and the blue luminosity measured from the HST image, this ratio is $0.5\pm0.2$, larger than the luminosity-weighted estimate but smaller than the mass-weighted estimate.

The $\log \,($\NII$/$\OII$)$ of the arc galaxy is $-0.4$, which meets the criterion for the upper $R_{23}$ branch \citep{Kewley2008}. Using the metallicity calibration of \citet{Zaritsky1994}, we obtained $12 + \log(\rm O/ \rm H)$ to be 8.88. Assuming a solar metallicity of $12 + \log(\rm O/ \rm H)=8.86$ \citep{Delahaye2006}, the metallicity is estimated to be $Z/Z_{\odot} \sim 1.0$, similar to the value derived from the STARLIGHT mass-weighted measurement and consistent with the value of $Z/Z_{\odot}=(0.3-1.3)$ estimated from 10 different mass-metallicity relations listed in \citet{Kewley2008}. These results are not sensitive to QSO contamination; consistent results are obtained without removing the QSO contamination.

\subsection{X-Ray spectra with \textit{XMM-Newton}}

\begin{deluxetable}{lll}
\tablecolumns{3}
\tablewidth{0pc}
\tablecaption{X-Ray spectral parameters of the QSO
\label{tab:xmm}}
\tablehead{
\colhead{Parameter} & \colhead{2009} & \colhead{2014}
}
\startdata
$N_{\rm H,Gal}$ ($10^{20}$~cm$^{-2}$) & 2.07 fixed & 2.07 fixed \\
$N_{\rm H,ext}$ ($10^{20}$~cm$^{-2}$) & $3.6_{-1.2}^{+1.4}$ & $1.608_{-0.009}^{+0.010}$ \\
PL photon index & $2.24_{-0.10}^{+0.12}$ & $1.96 \pm 0.08$ \\
PL norm\tablenotemark{a} & $2.13_{-0.19}^{+0.23}$ & $1.88_{-0.13}^{+0.15}$ \\
$E_{\rm line}$ (keV) & $2.12 \pm 0.03$ & \nodata \\
$A_{\rm line}$ ($10^{-6}~$ph~cm$^{-2}$~s$^{-1}$) & $-3.9 \pm 0.13$ & \nodata \\
EW (eV) & 71 & \nodata \\
$f_{\rm 0.3-10\;keV}$ ($10^{-13}~$\ergcms) & $4.04 \pm 0.24$ & $5.08 \pm 0.26$ \\
$f_{\rm 2-10\;keV}$ ($10^{-13}~$\ergcms) & $1.81 \pm 0.23$ & $2.70 \pm 0.25$ \\
$L_{\rm 0.3-10\;keV}$ ($10^{44}~$\ergs) & $3.01_{-0.19}^{+0.24}$ & $3.07 \pm 0.14$ \\
$L_{\rm 2-10\;keV}$ ($10^{44}~$\ergs) & $1.05 \pm 0.10$ &  $1.47 \pm 0.10$ \\
$\chi^2$ / degree of freedom & 96.5/93 & 135.9/121 
\enddata
\tablecomments{The XSPEC models are {\tt TBabs $\ast$ phabs(zgauss + zpowerlw)} for the 2009 observation and {\tt TBabs $\ast$ phabs $\ast$ zpowerlw} for the 2014 observation. $E_{\rm line}$ is the energy of the absorption feature in the rest frame. $f$ is the observed flux quoted in the observed frame and $L$ is the intrinsic luminosity corrected for absorption in the rest frame. All errors are quoted at the 90\% confidence level.}
\tablenotetext{a}{Power-law normalization in units of $10^{-4}$~photons~keV$^{-1}$~cm$^{-2}$~s$^{-1}$ at 1~keV.}
\end{deluxetable}


\begin{figure*}
\centering
\includegraphics[width=0.49\textwidth]{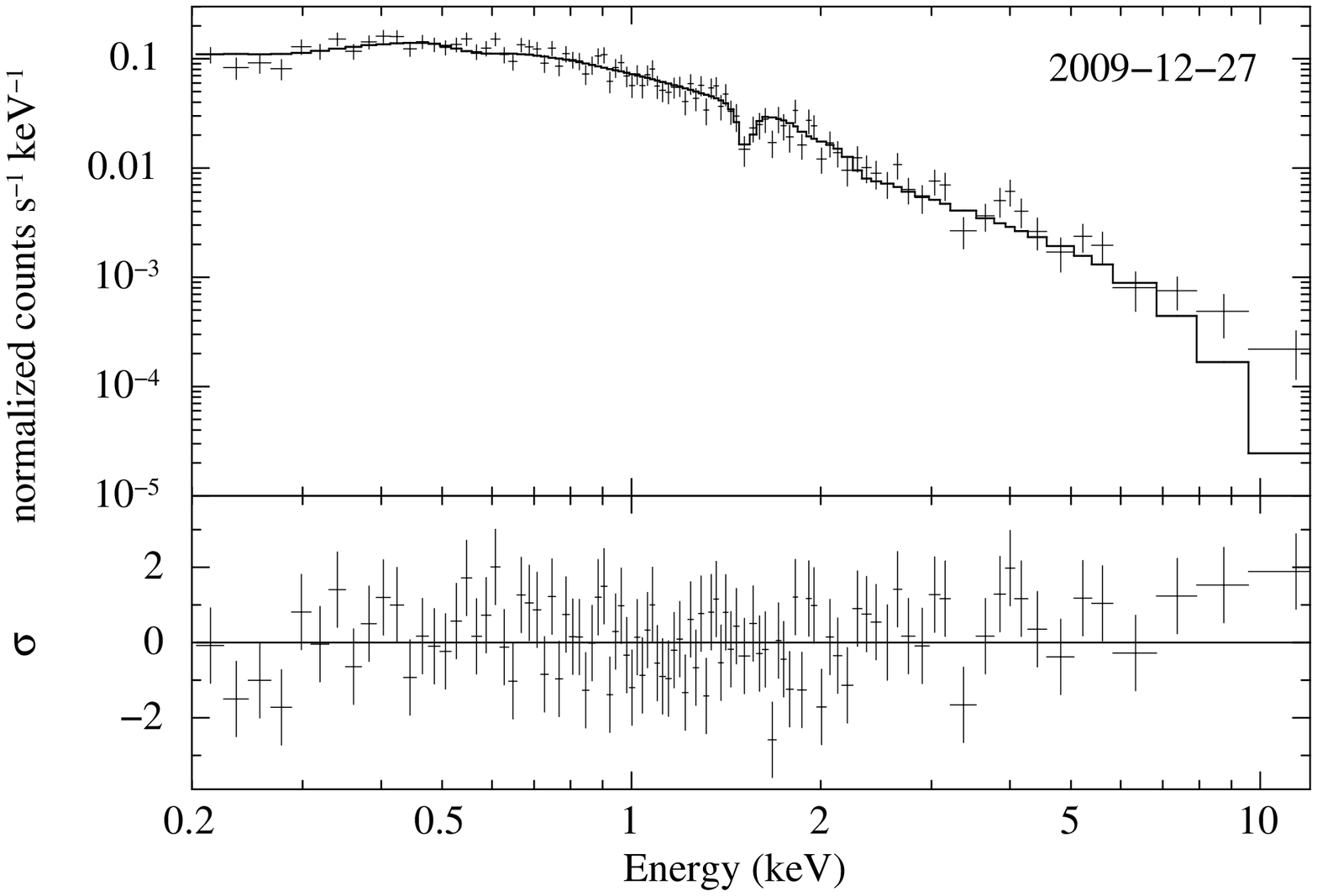}\hfill
\includegraphics[width=0.49\textwidth]{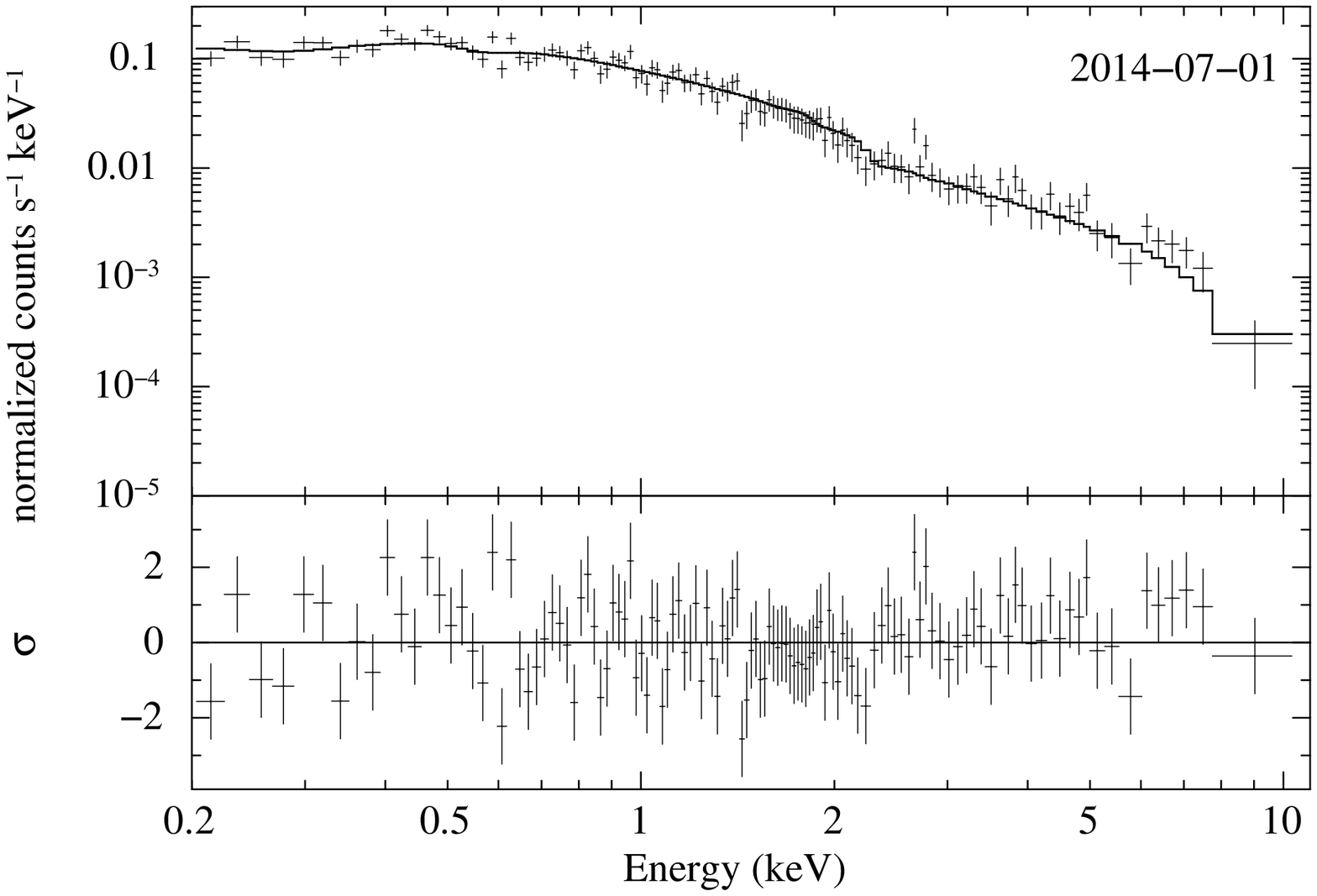}
\caption{XMM spectra of the QSO with best-fit models for the two observations; see the spectral parameters in Table~\ref{tab:xmm}. 
\label{fig:xmm}}
\end{figure*}

\begin{figure*}
\centering
\includegraphics[width=0.98\textwidth]{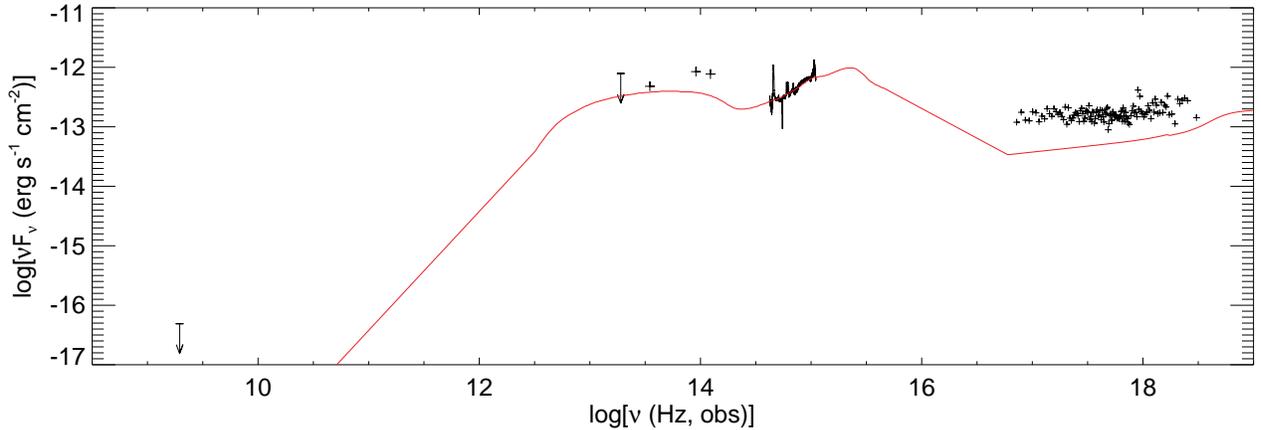}
\caption{Multiwavelength SED of the QSO from radio to X-ray energies in the rest frame. The red line indicates a QSO SED template from \citet{Hopkins2007}, normalized to match the optical spectrum.
\label{fig:sed}}
\end{figure*}

The {\it Chandra} observation suggests that the X-ray emission arises from a point-like source spatially coincident with the QSO; there is no X-ray emission detected in the arc region \citep{Tao2012}. {\it XMM-Newton} observed the galaxy on 2009 December 27 and 2014 July 1 (ObsID 0601010101 and 0728190101, respectively). An earlier observation made on 2001 July 8 was not used due to heavy background contamination. Only data from the PN CCD were used for analysis. New events files were created with up-to-date calibration files. Events were selected from low background intervals, where the background flux is within $\pm 3 \sigma$ of the mean quiescent level, adding up to an effective exposure of 19.0~ks and 24.3~ks, respectively, for the two observations. The source energy spectra were extracted from a circular region of 32\arcsec radius, and the background spectra were extracted from nearby circular regions on the same chip at a similar readout distance. The spectral bins were grouped such that each new bin is 1/4 of the local FWHM and has at least 15 counts, from 0.2 keV to 10 keV.

We tried to fit the energy spectra with a redshifted power-law model subject to interstellar absorption. The {\tt TBabs} model \citep{Wilms2000} is used to account for Galactic absorption, and the column density is fixed at the Galactic value $2.07 \times 10^{20}$~cm$^{-2}$ \citep{Kalberla2005}, while {\tt phabs} is adopted for additional extragalactic absorption. For the 2009 observation, we obtained consistent results with those reported by \citet{Jin2011}, if the same model and energy range are used. However, we found that a simple power-law model is insufficient to fit the data. An absorption feature near 2.12~keV in the rest frame (or 1.54 keV in the observed frame) and excessive emission above 10~keV in the rest frame (or 7 keV in the observed frame) are possibly shown in the residual. We further added a zero-width Gaussian component to fit the absorption feature.  The addition of the absorption line reduced the $\chi^2$ by 24.6, corresponding to a chance probability of $2.6\times10^{-5}$. The hard excess could be due to a reflection component, which is often seen in the spectra of AGNs and Galactic accreting black holes, but the current data quality and energy coverage do not allow us to quantify it. For the 2014 observation, a simple power-law model can adequately fit the data. The spectra are shown in Figure~\ref{fig:xmm} and the best-fit parameters are listed in Table~\ref{tab:xmm}.

\subsection{Multiwavelength SED}

NGC 247 was observed by the Wide-field Infrared Survey Explorer ({\it WISE}) in 2010 with the passbands W1 (3.4 $\mu$m), W2 (4.6 $\mu$m), W3 (12 $\mu$m) and W4 (22 $\mu$m). The QSO was detected in the W1, W2 and W3 bands with a signal-to-noise ratio (SNR) larger than 9, while in the W4 band, the source was not detected. From the AllWISE Source Catalog in NASA/IPAC Infrared Science Archive (IRSA)\footnote{http://irsa.ipac.caltech.edu/applications/Radar/}, we obtained the profile-fitting photometry for the W1, W2, and W3 bands in VEGA magnitudes, and a 95\% upper limit for the W4 band, which are $14.23\pm0.03$, $13.17\pm0.03$, $10.88\pm0.12$ and $<8.261$ for the W1, W2, W3 and W4 bands, respectively. Using the zero magnitude flux density and the color corrections from \citet{Wright2010}, the observed VEGA magnitudes were translated to flux density, which are $(6.26\pm0.18)\times10^{-27}$, $(9.3\pm0.3)\times10^{-27}$, $(1.37\pm0.15)\times10^{-26}$ and $ < 4.1\times10^{-26}$~\fnunu\ for the four bands in the same order. Moreover, the QSO was not detected in the 1.4 GHz NRAO VLA Sky Survey (NVSS), suggesting a flux less than 2.5 mJy \citep{Elvis1997}. The multiwavelength spectral energy distribution (SED) from the radio to the X-ray band with a QSO SED template renormalized in the optical band \citep{Hopkins2007} is shown in Figure~\ref{fig:sed}.

\section{Discussion}

The consistent redshifts of the two galaxies suggest that they are a merger event, between a luminous QSO (PHL 6625) and a tidally distorted companion galaxy, instead of a strong gravitational lens system. The stellar content of the two galaxies are not in contact yet, and the mass fraction of young stars is less than 1\% (Figure~\ref{fig:arc_sfh}), suggesting they are at the early stage of a merging process. Such a system seems to be an analogue of the nearby event Arp 142 (NGC 2936/37) in morphology \citep[e.g.,][]{Romano2008}, except that the central black hole in PHL 6625 is an active quasar.

\subsection{A major or minor merger?}

The broadening of the QSO image (Figure~\ref{fig:profile}) suggests that an underlying component possibly due to its host galaxy is detected. However, due to the saturation of the central pixels, the galaxy bulge cannot be spatially resolved. A conservative estimate of the $B$-band luminosity of the QSO host galaxy is $(1.00 - 6.93) \times 10^{10}$~$L_\sun$, corresponding to a mass of $(4-28) \times 10^{10}$~$M_\sun$ assuming a typical mass-to-light ratio of 4 \citep{Faber1979}. The black hole mass for the QSO is estimated to be $(2-5) \times 10^8$~$M_\sun$ via different techniques. Assuming an $M_{\rm BH}$-$M_{\rm bulge}$ relation \citep{Kormendy2013},  
\begin{equation}
\frac{M_{\rm BH}}{10^{9} \; M_\odot} = 
0.49
\left( \frac{M_{\rm bulge}}{10^{11} \; M_\odot} \right)^{1.17}, 
\end{equation} 
we can derive the bulge mass to be $(4-11) \times 10^{10}$~$M_\sun$ with an intrinsic scatter of 0.28~dex (a factor of $\sim$2). This is consistent with the mass range estimated from image decomposition, and suggests that the QSO host galaxy may be an elliptical or a bulge-dominated system. If we adopt the total mass range of the host galaxy, PHL 6625 also follows the distribution of AGNs at $z=0.1-1.0$ in the $M_{\rm BH}$-$M_{\rm bulge}$ plane \citep[Figure 38 of][]{Kormendy2013}, and is consistent with the $M_{\rm BH}$-$M_{\rm bulge}$ relation at $z=0.4$ \citep{Kormendy2013}. In summary, the QSO and its host seem to be a canonical example on the co-evolution path.

The stellar mass of the arc galaxy, using the population synthesis measurement, is about $6.8 \times 10^{9}$~$M_\sun$, indicating that the mass ratio of the QSO host galaxy and the arc galaxy is around 10,  suggesting that the system is a minor merger. Given the $B$-band luminosity of the arc galaxy of $1.4 \times 10^{10}$ $L_\sun$ and $M_\ast/L_{\rm B} \sim 0.1-0.2$ derived from the luminosity-weighted metallicity of the population synthesis model, the stellar mass of the arc galaxy is on the order of $10^{9}$~$M_\sun$, and the system is also likely to be a minor merger. But if $M_\ast/L_{\rm B}$ derived from the mass-weighted metallicity is used, the stellar mass of the arc galaxy will be more than 10 times higher, and the system could be a major merger. However, during the population synthesis fit, the mass-to-light ratio of young stars (age $\sim 10^{8.5}$~year) is about one-sixth that of old stars (age $\sim 10^{10}$~year), then any small light-fraction variations of young stellar populations will make large mass-fraction variations on old stellar populations and result in huge uncertainties in the mass-weighted metallicity and mass-to-light ratio. Thus, the luminosity-weighted measurement is more reliable than the mass-weighted measurement, and the system is more likely to be a minor merger, although a major merger cannot be excluded.

\subsection{Multiwavelength properties of the QSO}

While there is some weak evidence for mild spectral variability, the QSO outputs a consistent luminosity in the X-ray band ($\sim 3\times10^{44}$~\ergs\ in 0.3-10~keV, rest frame). The bolometric luminosities, $L_{\rm bol}$, calculated from the unabsorbed flux in the 0.01--100 keV range, are, respectively, $1.1\times10^{45}$~\ergs\ and $8.0\times10^{44}$~\ergs\ for the 2009 and 2014 observations, corresponding to an Eddington ratios ($L_{\rm bol}/L_{\rm Edd}$) of about $0.01-0.05$. This implies a bolometric correction factor of 3--5 for the $L_{5100}$ luminosity or 5--11 for the 2-10 keV X-ray luminosity. 

The observed power-law photon index and the Eddington ratio are consistent with the $\Gamma - L_{\rm bol}/L_{\rm Edd}$ relation for AGNs \citep{Brightman2013}. The hardening of the spectrum along with the decrease of the bolometric luminosity between the 2009 to 2014 observations, if true, is also consistent with the above relation. 

The X-ray to optical/UV ratio, $\alpha_{\rm ox}$, is defined as
\begin{equation}
{\alpha_{\rm ox}} = 0.3838\log(L_{\rm 2~keV}/L_{\rm 2500~\angstrom}),
\end{equation} 
where $L_{\rm 2~keV}$ and $L_{\rm 2500~\angstrom}$ are the monochromatic luminosities at 2~keV and 2500~\AA\ in the rest frame, respectively. With $L_{\rm 2500~\angstrom}=4.68 \times 10^{29}$~\fnu\ and $L_{\rm 2~keV}=1.82 \times 10^{26}$~\fnu\ for the 2014 {\it XMM-Newton} observation, we derive $\alpha_{\rm ox} = -1.3$, which is consistent with the $\alpha_{\rm ox} = -1.4\pm0.3$ derived from the $\alpha_{\rm ox} - L_{\rm 2500~\angstrom}$ relation of \citet{Just2007}.

The \FeII\ strength of the QSO, defined as the ratio of the equivalent width for the optical \FeII\,$\lambda$4570 blend and the broad \hbeta\ ($R_{\rm Fe\,{\sc II}} \equiv {\rm EW_{Fe\,{\sc II}}}/{\rm EW_{H\beta}}$), is $\sim 0.14$. The FWHM of the broad \hbeta\ is 6447~km~s$^{-1}$. The measured log[EW$_{\rm [O\,{\sc III}]\,\lambda 5007}$~(\AA)] is about 0.8. These properties are not typical for SDSS quasars \citep{Shen2014} on the eigenvector 1 plane \citep{Boroson1992}, although the QSO presents a typical luminosity among the SDSS DR7 quasars \citep{Shen2011}. Quasars similar to PHL 6625 that have a small $R_{\rm Fe\,{\sc II}}$ and a low equivalent width of \OIIIb\ compose only 3.4\% of the SDSS DR7 quasars with a small $R_{\rm Fe\,{\sc II}}$, regardless of \OIIIb\ strength, but the physical explanation is not clear.

To conclude, this system gives us a case where a luminous quasar is associated with a minor merger in the close pair phase, although there is no conclusive evidence to show a link between the quasar activity and the merger event. High-resolution and high-sensitivity observations with a large sample of nearby quasars may address the question whether the case like PHL 6625 is rare or ubiquitous, and whether quasar activity can be triggered by minor mergers.

\acknowledgements 
We thank the anonymous referee for useful comments that have improved the paper. We also thank Chien Y. Peng, Minjin Kim, Ning Jiang, and Yulin Zhao for their help in using of the GALFIT software. H.F. acknowledges funding support from the National Natural Science Foundation of China under grant No.\ 11633003, and  the National Program on Key Research and Development Project (grant No. 2016YFA040080X). L.C.H. was supported by the National Key Program for Science and Technology Research and Development grant 2016YFA0400702. J.Q.G. and S.M. were partially supported by the Strategic Priority Research Program ``The Emergence of Cosmological Structures'' of the Chinese Academy of Sciences grant No.\ XDB09000000 and by the National Natural Science Foundation of China (NSFC) under grant numbers 11333003 and 11390372.

{\it Facilities:} \facility{VLT}, \facility{HST}, \facility{XMM-Newton}

\end{document}